\documentclass[pss,fleqn]{w-art}
\usepackage{times}
\usepackage{w-thm}
\usepackage[]{graphicx}
\begin{document}
\newcommand {\bea}{\begin{eqnarray}}
\newcommand {\eea}{\end{eqnarray}}
\newcommand {\bb}{\bibitem}
%%    The information for the title page will be placed between
%%    \begin{document} and \maketitle. The order of most entries
%%    is determined by the class file and can not be changed by
%%    rearranging them. The maketitle command follows after the
%%    abstract.
%%
%%    Most of the following commands will be completed by the publisher.
%%
%%    The copyrightyear is defined in the .clo file as the first argument
%%    of the copyrightinfo command. If the copyrightyear differs from that
%%    value it might be adjusted by the following definition:
%%
%% \renewcommand{\copyrightyear}{2003}% uncomment to change the copyrightyear.
%%
\DOIsuffix{theDOIsuffix}
%%
%% issueinfo for header and copyright line
\Volume{XX}
\Issue{1}
\Month{01}
\Year{2003}
%%
%%    First and last pagenumber of the article. If the option
%%    'autolastpage' is set (default) the second argument may be left empty.
\pagespan{1}{}
%%
%%    Dates will be filled in by the publisher. The 'reviseddate' and
%%    'dateposted' (Published online) entry may be left empty.
\Receiveddate{}
\Reviseddate{}
\Accepteddate{}
\Dateposted{}
\title{Gossamer Superconductivity, New Paradigm?}

%% Please do not enter footnotes or \inst{}-notes into the optional
%% argument of the author command. The optional argument will go into
%% the header.  If there is only one address the marker \inst{x} may be
%% omitted.

%% Information for the first author.
\author[H. Won]{Hyekyung Won\footnote{Corresponding
     author: e-mail: {hkwon@hallym.ac.kr}}\inst{1}} \address[\inst{1}]
{Department of Physics, Hallym University, Chuncheon 200-702, South Korea}
%%
%%    Information for the second author
\author[S. Haas]{Stephan Haas\inst{2}}
\address[\inst{2}]{Department of Physics and Astronomy, University of Southern
California, Los Angeles, CA 90089-0484 USA}
%%
%%    Information for the third author
\author[K. Maki]{Kazumi Maki\inst{2,3}}
\address[\inst{3}]{Max-Planck Institute for the Physics of Complex Systems,
N\"{o}thnitzer Str. 38, D-01187, Dresden, Germany}

\author[D. Parker]{David Parker\inst{2}}

\author[B. Dora]{Balazs Dora\inst{4}}
\address[\inst{4}]{Department of Physics, Budapest University of Technology
and Economics, H-1521, Budapest, Hungary}

\author[A. Virosztek]{Attila Virosztek\inst{4,5}}
\address[\inst{5}]{Research Institute for Solid State Physics and Optics,
P.O. Box 49, H-1525 Budapest, Hungary}

\begin{abstract}
We shall review our recent works on d-wave density wave (dDW) and gossamer
superconductivity (i.e. d-wave superconductivity in the presence of dDW) in
high-T$_{c}$ cuprates and CeCoIn$_{5}$. a) We show that both the giant
Nernst effect and the angle dependent magnetoresistance (ADMR) in the 
pseudogap phases of the cuprates and CeCoIn$_{5}$ are manifestations of dDW.
b) The phase diagram of high-T$_{c}$ cuprates is understood in terms of
mean field theory, which includes two order parameters $\Delta_{1}$ and
$\Delta_{2}$, where one order paremeter is from dDW and the other from d-wave
superconductivity. c) In the optimally to the overdoped region we find 
the spatially periodic dDW, an analogue of the Fulde-Ferrell-Larkin-Ovchinnikov
(FFLO) state, becomes more stable. d) In the underdoped region where
$\Delta_{2}/\Delta_{1} \ll 1$ the Uemera relation is obtained within the 
present model.  We speculate that the gossamer superconductivity is at the 
heart of high-T$_{c}$ cuprate superconductors, the heavy-fermion 
superconductor CeCoIn$_{5}$ and the organic superconductors $\kappa$-
(ET)$_{2}$Cu(NCS)$_{2}$ and (TMTSF)$_2$PF$_{6}$.
\end{abstract}
\maketitle                   % Produces the title.

%% If there is not enough space inside the running head
%% for all authors including the title you may provide
%% the leftmark in one of the following three forms:

%% \renewcommand{\leftmark}
%% {First Author: A Short Title}

%% \renewcommand{\leftmark}
%% {First Author and Second Author: A Short Title}

%% \renewcommand{\leftmark}
%% {First Author et al.: A Short Title}

%% \tableofcontents  % Produces the table of contents.
\section{Introduction}

\textbf{Question:} What does strong correlation mean? \\
\textbf{Answer:} First of all it means the Coulomb dominance; the Coulomb
interaction is stronger than that due to phonon exchange.  For superconductors
this means unconventional order parameters: d-wave,f-wave,
g-wave superconductors.
\\
\\
Since the discovery of high-T$_{c}$ cuprates La$_{2-x}$Ba$_x$CuO$_{4}$ by
Bednorz and M\"{u}ller \cite{1} in 1986, it appears that the debate over
the nature and mechanism of this unusual superconductivity continues.
However, d-wave superconductivity as in BCS theory and arising due to 
anti-paramagnon exchange has been established, at least in the vicinity of 
the optimal doping.\cite{2,3}  Also from the low temperature thermal
conductivity May Chiao et al deduced $\Delta/E_{F} = 1/10$ and $1/14$ for
Bi-2212 and YBCO respectively \cite{4,5}.  From these we obtain $\Delta= 500 K$
and $280 K$ for Bi-2212 and YBCO respectively and E$_{F} \simeq 5000 K$, which
is almost universal.  Here $\Delta$ is the maximal gap of d-wave
superconductivity at $T=0$.  Also recently the universality of the
Fermi velocity $v = 2.3 \times 10^{7}$ cm/sec has been established by the
angle resolved photoemission spectrum \cite{6}.

From these we conclude that the cuprate superconductors are in the
BCS limit, far away from the Bose-Einstein condensation limit,
and that the superconducting fluctuation effect should
be at most 10\%.  Therefore theories based on the large superconducting
fluctuations \cite{7,8} appear to be unrealistic.  Also in many numerical
computations on the cuprates, it was assumed that $\Delta \sim E_{F}$.
It is clear that such approximations lead to rather unrealistic 
predictions.  On the contrary, with $\Delta=0.1 E_{F}$ Kato et al recently
found hundreds of quasiparticle
bound states around a vortex of f-wave superconductors, a model system
for Sr$_{2}$RuO$_{4}$ \cite{9}.  These bound states are the analogues of 
Caroli, de Gennes and Matricon bound states around a vortex in s-wave 
superconductivity \cite{10,11}.  More recently the STM data around a 
vortex in Sr$_{2}$RuO$_{4}$ has been reported by Lupien et al \cite{12}.
Indeed the observed quasiparticle spectrum is very consistent with the
theoretical analysis in \cite{9}.  Of course in Sr$_{2}$RuO$_{4}$ $\Delta$
should be less than 0.01 E$_{F}$.

In 1993 Volovik \cite{13} showed that the quasiparticle density of states
in the vortex state in d-wave superconductors is calculable within a 
quasiclassical approximation.  This work has been extended into several
directions: a) thermodynamic functions; b) thermal conductivity; c) scaling
relations; d) for arbitrary field orientation; and e) for a variety of gap
functions $\Delta({\bf k})'s$ \cite{14,15,16,17,18,19}.  As is well known
the gap symmetry and the gap function has been the central issue since
the discovery of the heavy-fermion superconductors \cite{20}.  Since 2001
Izawa et al have succeeded in determining the superconducting gap functions
$\Delta({\bf k})'s$ in Sr$_{2}$RuO$_{4}$ \cite{21}, CeCoIn$_{5}$ \cite{22},
$\kappa$-(ET)$_{2}$Cu(NCS)$_{2}$ \cite{23}, YNi$_{2}$B$_{2}$C \cite{24},
PrOs$_{4}$Sb$_{12}$ \cite{25,26}, UPd$_{2}$Al$_{3}$ \cite{27,28} and
CePt$_{3}$Si \cite{29,30} through the angle-dependent thermal conductivity.
For a review of these aspects see \cite{31}.

The phase diagrams of high-T$_{c}$ cuprates and CeCoIn$_{5}$ are shown
in Fig. 1a) and Fig. 1b) respectively.  As you may recognize, we have replaced
\begin{figure}[h]
\includegraphics[width=7.2cm]{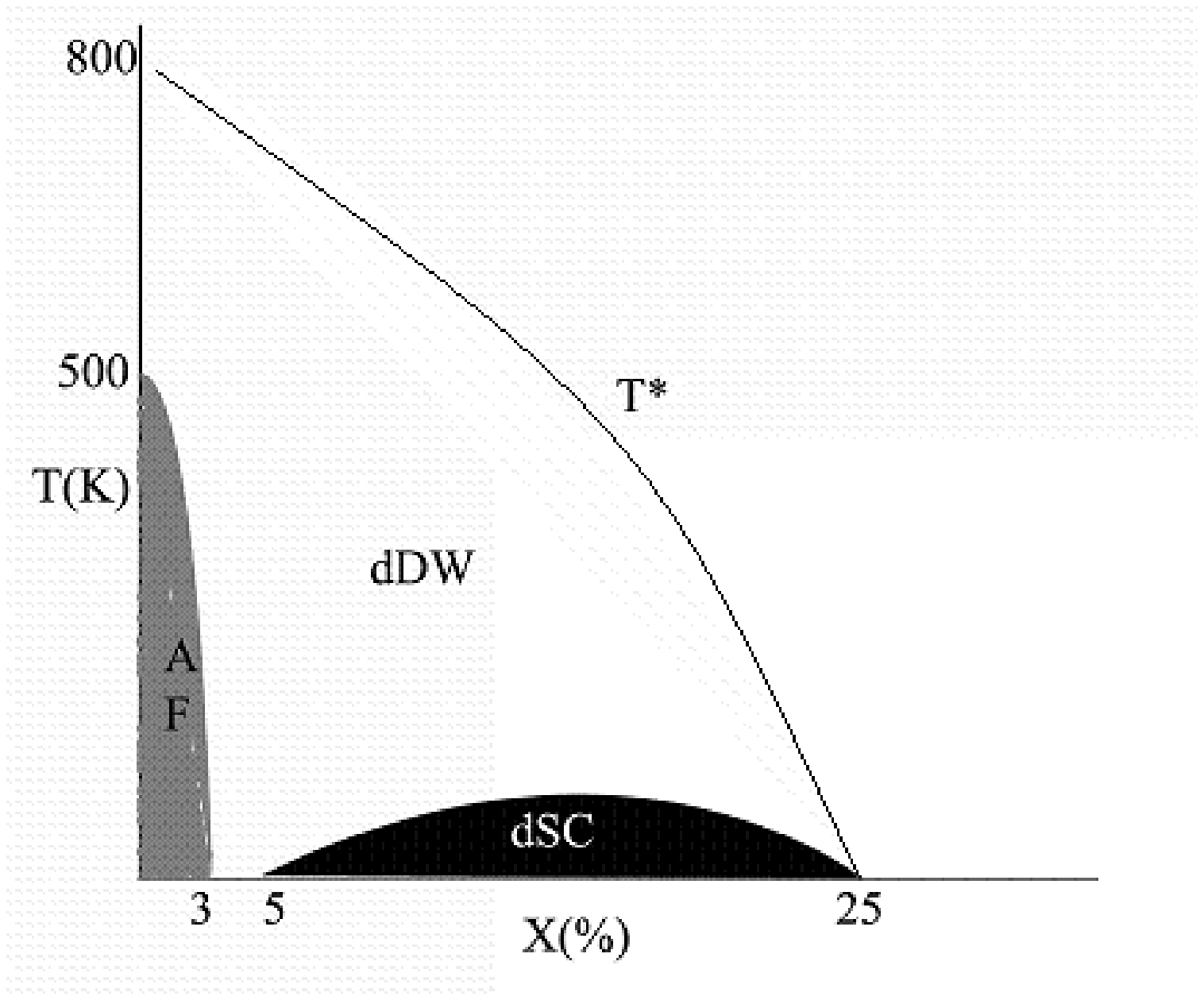}
\includegraphics[width=7.2cm]{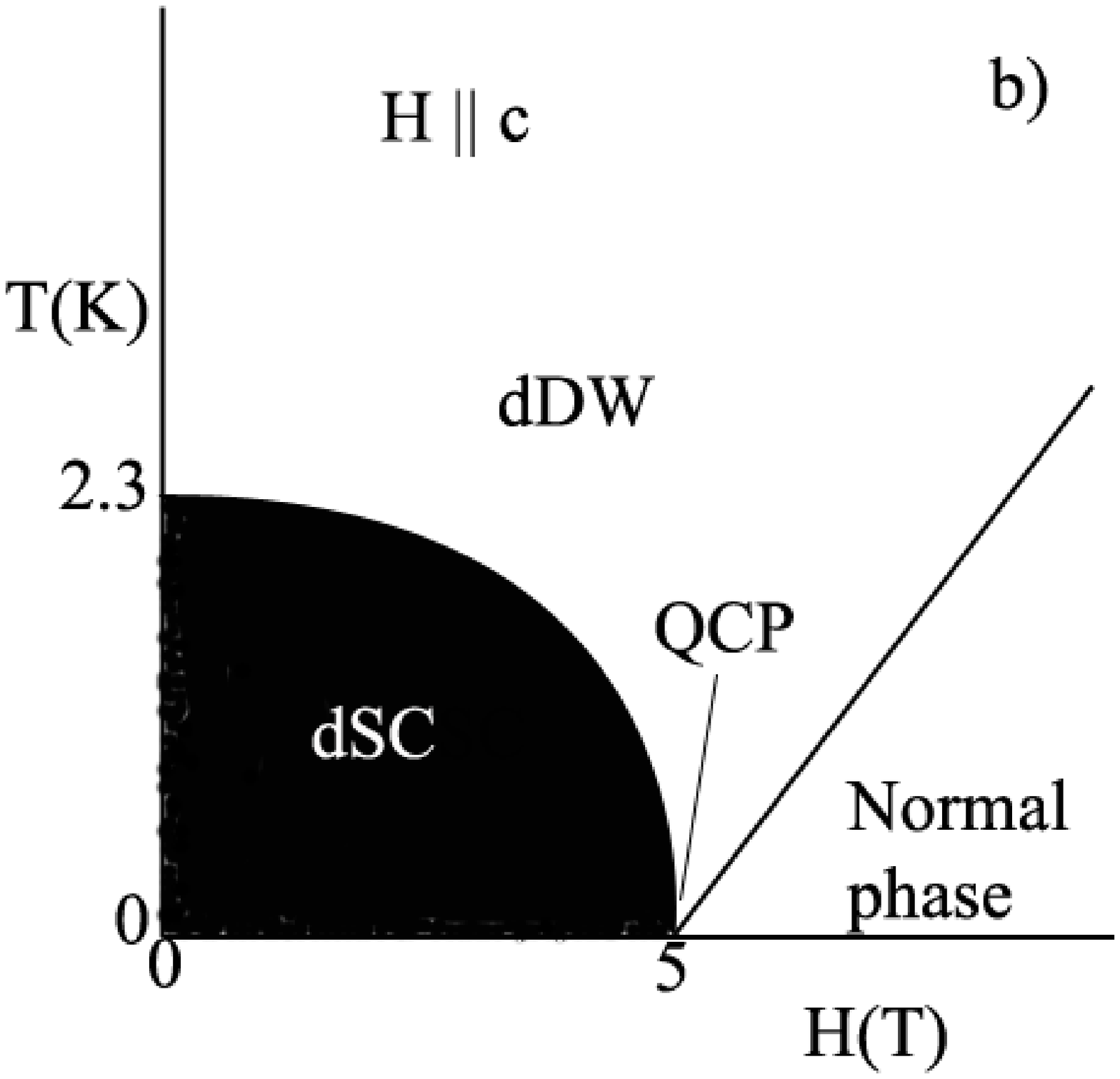}
\caption{The phase diagrams for high-T$_{c}$ cuprates (left) and CeCoIn$_{5}$}
\end{figure}
the pseudogap phase with d-wave density wave, 
which is the main topic in section 2.
Also we have chosen T* to vanish at the same point where the superconductivity
vanishes, suggesting the system has a quantum critical point at $x=25\%$ 
\cite{32}.  D-wave density wave (dDW) for the pseudogap phase in high-T$_{c}$
cuprates has been proposed by several people. \cite{33,34,35,36}.  However,
unlike these authors we do not consider the commensurate dDW with Z$_{2}$-
symmetry, which is a descendant of the flux phase \cite{37}.  Rather we
limit ourselves to the incommensurate dDW with the U(1) gauge symmetry as in
the conventional charge density wave \cite{38,39}.  

We shall see later that the incommensurate dDW is crucial to understanding
the phase diagram in Fig. 1a).  Also we shall discuss the recent angle 
dependent magnetoresistance (ADMR) data in Y$_{0.68}$Pr$_{0.32}$CuO$_{4}$
\cite{40,41} and CeCoIn$_{5}$ \cite{42}, which provides strong support
for dDW in these systems.  Once one accepts the phase diagrams in Fig. 1, the
d-wave superconductivity in the high-T$_{c}$ cuprates and CeCoIn$_{5}$ should
arise in the presence of dDW.  Borrowing the beautiful word from R.B. Laughlin
\cite{43} we call these superconductors ``gossamer superconductors''.  
Therefore the exploration of the gossamer superconductivity appears
to be the most urgent \cite{44}.  We shall interpret the Uemura relation
in the vicinity of $x=5\%$ in terms of this gossamer superconductivity.

\section{D-wave density waves}

There are many parallels between the cuprates, the heavy-fermion superconductor
CeCoIn$_{5}$ and the organic conductor $\kappa$-(ET)$_{2}$Cu(NCS)$_2$:
the quasi-two dimensional Fermi surface, the proximity of the
antiferromagnetic phase and d-wave superconductivity \cite{22,45}.  In addition
d-wave density wave in the pseudogap phase appears to be an additional
common feature \cite{41,42,46,47}.  In the absence of a magnetic field the
Nambu Green function for dDW is given by \cite{41}
\bea
G^{-1}(\omega,{\bf k}) &=& \omega - \xi({\bf k})\rho_{3} - \eta({\bf k}) -
\Delta({\bf k})\rho_{1}
\eea
where the $\rho_{i}$'s are the Pauli matrices operating on the spinor space.
For d-wave charge density wave we can take either $\Delta({\bf k}) =
\Delta\cos(2\phi)$ or $\sin(2\phi)$ with $\tan\phi = k_{y}/k_{x}$ and
$\eta({\bf k})= \mu$, the chemical potential, which acts as the imperfect
nesting.  Further
\bea
\xi({\bf k}) = v(k_{\parallel}-k_{F})+\frac{v^{'}}{c}\cos(ck_{z})
\eea
where $k_{\parallel}$ is the radial component in the x-y plane and
v and v$^{'}$ are the Fermi velocities.

Then the quasiparticle density of states is given by
\bea
N(E)/N_{0} &=& G(x-y)
\eea
where 
\bea
G(x)&=& \frac{2x}{\pi}K(x)\,\,\,\mathrm{for\,\, x \leq 1} \\
    &=& \frac{2}{\pi}K(x^{-1})\,\,\,\mathrm{for\,\, x > 1}.
\eea
\begin{figure}[h!]
\includegraphics[width=7cm,angle=270]{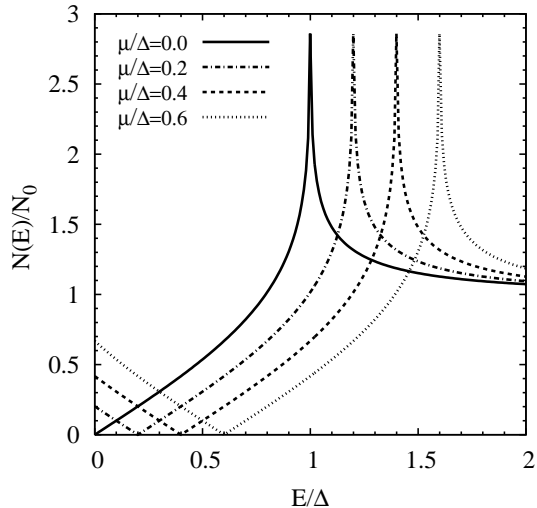}
\caption{The quasiparticle density of states for a dDW superconductor}
\end{figure}
and $x=E/\Delta$, $y= \mu/\Delta$ and K(x) is the complete elliptic
integral.  $N(E)/N_{0}$ is shown in Fig. 2.  

Note that $N(0)/N_{0} \simeq |\mu|/\Delta$ for $\mu < \Delta$.
Therefore the chemical potential provides nonvanishing quasiparticle
density of states at E=0.  This gives rise to pockets on the Fermi surface
at $(\pi,\pi)$ direction in the high-T$_{c}$ 
cuprates \cite{48}.  Also as we shall see
later $\Delta_{2} \leq \mu$ is crucial for the presence of d-wave
superconductivity in the middle of dDW.  Here $\Delta_{2}$ is the maximum
energy gap of a d-wave superconductor.

\section{Landau Quantization}

As noted by Nersesyan et al \cite{49,50} the quasiparticle spectrum
is quantized in the presence of a magnetic field.  Let us consider a
magnetic field ${\bf B}$ applied within the $x^{'}-z$ plane tilted
by an angle $\theta$ from the z axis.  Also $\hat{x}^{'}$ is defined
by $\hat{x}^{'}=\hat{x} \cos\phi + \hat{y}\sin\phi$.  Then the magnetic 
field is introduced by ${\bf k} \rightarrow {\bf k}+e{\bf A}$ with
\bea
{\bf A} &=& B(y\cos\phi-x\sin\phi)(\hat{z}\sin\theta+(\hat{x}\cos\phi+
\hat{y}\sin\phi)\cos\theta)
\eea
Then for d$_{xy}$-wave DW, the quasiparticle energies are given by
\bea
E^{\pm}_{1n} &=& \pm \sqrt{2neBv_{2}|(v\cos\theta\cos\phi-v^{'}\sin\theta)
\cos\phi|} - \mu \\
E^{\pm}_{2n} &=& \pm \sqrt{2neBv_{2}|(v\cos\theta\cos\phi+v^{'}\sin\theta)
\cos\phi|} - \mu \\
E^{\pm}_{3n} &=& \pm \sqrt{2neBv_{2}|(v\cos\theta\sin\phi-v^{'}\sin\theta)
\sin\phi|} - \mu \\
E^{\pm}_{4n} &=& \pm \sqrt{2neBv_{2}|(v\cos\theta\sin\phi+v^{'}\sin\theta)
\sin\phi|} - \mu 
\eea
Here n=0,1,2,etc.  Except for the n=0 Landau level they are doubly
degenerate.  Also unlike the quasi-one dimensional systems \cite{39},
there are 4 branches of the Landau levels \cite{41}.  As shown elsewhere
these Landau spectra are most readily seen by angle dependent magnetoresistance
(ADMR), the nonlinear Hall conductivity and the giant Nernst effect 
\cite{39,51}.  Indeed ADMR appears to provide the most sensitive test
of unconventional density wave (UDW) as seen in 
$\alpha$-(ET)$_{2}$KHg(SCN)$_{4}$ and the Bechgaard salts (TMTSF)$_{2}$X
with X=PF$_{6}$ and ReO$_{4}$ \cite{52,53,54}.  Here we present such
an analysis of ADMR data provided by C. Almasan, T. Hu and 
V. Sandu in the pseudogap region in Y$_{0.68}$Pr$_{0.32}$CuO$_{4}$
\cite{40,41} and CeCoIn$_{5}$, which are shown in Fig. 3a and 3b.
\begin{figure}[h]
\includegraphics[width=8cm]{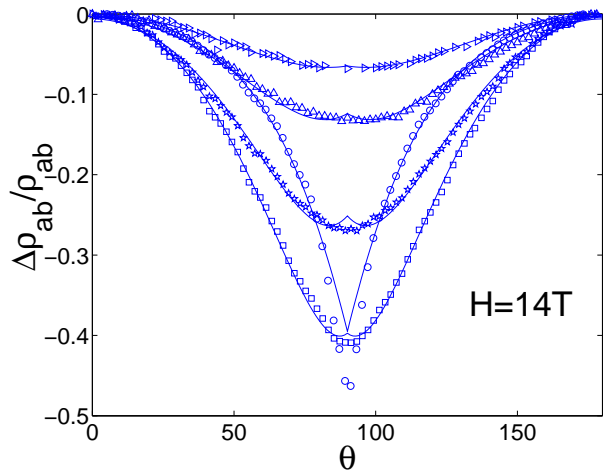}
\caption{ADMR of pseudogap region of Y$_{0.68}$Pr$_{0.32}$CuO$_{4}$.
Curves are for T= 105 K, 75 K, 65 K, 60 K and 52 K from top to bottom.  
The curve for T =52 K is reduced by a factor of 10.}
\end{figure}
\begin{figure}[h]
\includegraphics[width=8cm]{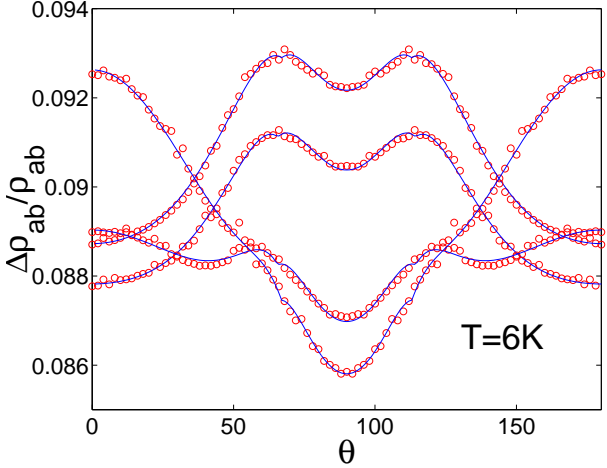} 
\caption{ADMR of pseudogap region of CeCoIn$_{5}$, for H = 3 T,
5 T, 8 T and 14 T from top to bottom \cite{42}.}
\end{figure}The electric conductivity is given by \cite{41}
\bea
\sigma(B,\theta) =\sum_{n}\sigma_{n} \mathrm{sech^{2}}(\beta E_{n}/2) 
\eea
where $\beta= 1/k_{B}T$ and the sum is over all the Landau lavels.
However, when $\beta |E_{1}| \gg 1$, the two lowest Landau levels
suffice.  Then we obtain 
\bea
\sigma(B,\theta) &=& \sigma_{0}^{'}(1+\cosh\zeta_{0})^{-1}+\sigma_{1}^{'}
\left[\frac{1+\cosh x_{1}\cosh \zeta_{0}}{\cosh x_{1}+\cosh \zeta_{0}} + 
\frac{1+\cosh x_{2}\cosh \zeta_{0}}{\cosh x_{2}+\cosh \zeta_{0}}\right]
\eea
where $\zeta_{0}=\beta \mu, x_{1} = \beta \sqrt{2eBv_{2}|v\cos\theta-v^{'}
\sin\theta|}$ and $ x_{2} = \beta \sqrt{2eBv_{2}|v\cos\theta+v^{'}\sin
\theta|}$.  Here we consider the case $\phi=0$ and $\sigma_{0}^{'}$ and
$\sigma_{1}^{'}$ are $\sigma_{0}$ and $\sigma_{1}$ multiplied by some
integer which accounts for the proper degeneracy.  From the fitting of
Fig. 3a) we obtain $v=2.3 \times 10^{7}$ cm/s, $v^{'}/v \leq 0.1$,
$E_{F} = 5000 K$, $\Delta = 360 K$ and $\mu \simeq 40 - 60 K$ for
Y$_{0.68}$Pr$_{0.32}$CuO$_{4}$ with T$_{c}$ = 55 K.  Similarly the data
from CeCoIn$_{5}$ is analyzed in \cite{42}.  We find $v=3.3 \times 10^{6} cm/s$, $v^{'}/v \simeq 0.5$,
$E_{F} = 500 K$, $\Delta = 45 K$ and $\mu = 8.4 K$ for CeCoIn$_{5}$.  These
values are consistent with other observations in CeCoIn$_{5}$ \cite{55}.
The Hall conductivity is given similarly by
\bea
\sigma_{xy} &=& - \frac{2e^{2}\cos^{2}\theta}{\pi}n(B,T)
\eea
with
\bea
n(B,T) &=& \tanh(\zeta_{0}/2) +\frac{\sinh(\zeta_{0})}{\cosh x_{1}+ 
\cosh \zeta_{0}} + \frac{\sinh(\zeta_{0})}{\cosh x_{2}+ 
\cosh \zeta_{0}} + \ldots
\eea
A similar expression has been obtained in \cite{54}.
The giant Nernst effect in the pseudogap phase of the high-T$_{c}$ cuprates
and CeCoIn$_{5}$ has already been discussed in \cite{46,47,56}.  In 
conclusion, ADMR, the non-linear Hall conductivity and the giant Nernst effect
should provide a clear signature of UDW.

\section{Gossamer Superconductivity}

Let us consider a simplest coupled equation for $\Delta_{1}$ (dDW) and 
$\Delta_{2}$ (d-wave superconductivity) \cite{44}:
\bea
\lambda_{1}^{-1} &=& 4\pi T \sum_{n} Re \left\langle 
\frac{f^{2}}{[(\sqrt{\omega_{n}^{2}
+\Delta_{2}^{2}f^{2}}-i\mu)^{2}+\Delta_{1}^{2}f^{2}]^{1/2}}\right\rangle \\
\lambda_{2}^{-1} &=& 4\pi T \sum_{n} Re \left\langle \frac{f^{2}(1-\frac{i\mu}
{\sqrt{\omega_{n}^{2}+\Delta_{2}^{2}f^{2}}})}{[(\sqrt{\omega_{n}^{2}
+\Delta_{2}^{2}f^{2}}-i\mu)^{2}+\Delta_{1}^{2}f^{2}]^{1/2}} \right \rangle
\eea
where $\lambda_{1}$ and $\lambda_{2}$ are dimensionless coupling constants,
f=$\cos(2\phi)$ and $\langle \ldots \rangle$ means $\int_{0}^{2\pi}\frac{d\phi}
{2\pi}$.  A similar set of equations is considered in \cite{36,57}.  First
let us consider Eq. (14) for $\Delta_{2} = 0 $.  Then we discover that the
equation is the same as for a d-wave superconductor in the presence
of the Pauli term \cite{58}.  Now if one puts $\Delta_{1} =0$, we obtain
the equation for $T_{c1}$ for dDW (=$T^*$) as
\bea \ln(\frac{T_{c1}}{T_{c10}}) &=& Re \Psi(\frac{1}{2}-\frac{i\mu}{2\pi 
T_{c1}})-\Psi(\frac{1}{2})
\eea
This is shown in Fig. 5.  Here $\Psi(z)$ is the digamma function and Eq.(17)
\begin{figure}[h]
\includegraphics[width=6cm,angle=270]{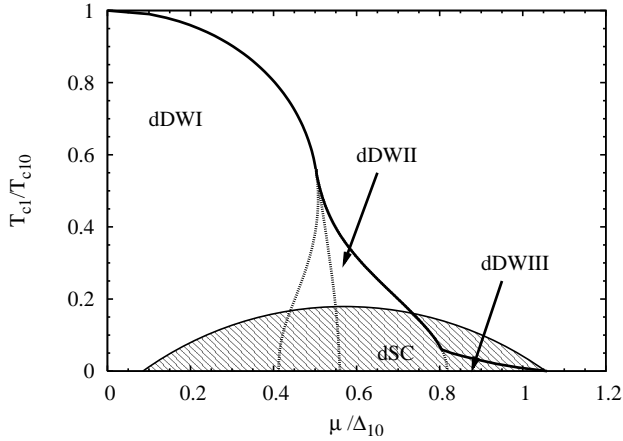}
\caption{Proposed phase diagram for high-T$_{c}$ cuprates}
\end{figure}
is the same as in s-wave superconductors \cite{59,60}.  The figure
for T$_{c1}$ bends back and there will be no solution for 
$\mu/\Delta_{10} > 0.57$.  On the other hand in the region
$0.41 < \mu/\Delta_{10} < 0.57$ T$_{c1}$ is double-valued.  \cite{58}.  
A similar phase
boundary is obtained numerically in \cite{36}.  However, the phase boundary
for dDW is extended if we allow the spatial variation of $\Delta$: 
$\Delta({\bf r}) \propto \cos({\bf q} \cdot {\bf r})$.

This is similar to the Fulde-Ferrell-Larkin-Ovchinnikov (FFLO) state in
d-wave superconductors \cite{61}, with T$_{c1}$ given by
\bea
-\ln(\frac{T_{c1}}{T_{c10}}) &=& Re \langle ( 1 \pm \cos(2\phi))
\Psi(\frac{1}{2} - \frac{i\mu(1-p\cos\phi)}{2\pi T_{c1}}) \rangle
- \Psi(\frac{1}{2})
\eea
where p is the new adjustable parameter.  Then $\pm$ in Eq. (18) corresponds
to ${\bf q} \parallel [100]$ and ${\bf q} \parallel [110]$ and $p= 
\frac{v|q|}{2\mu}$.  The extended solution is shown in Fig. 5 as well.  Then
it is more appropriate to split dDW in three separate regions as indicated
in Fig. 5.  Here we took $\Delta_{10}$ = 1700 K and T$_{c1}$ = 800 K in
accordance with Ref.\cite{65}.  
Finally we indicate the d-wave superconducting region by a 
shaded area, which should follow from the set of equations (15) and (16).
Also it is possible that dDW III may be submerged under the d-wave
superconductivity.

Since $T_{c1} > T_{c2}$ in general, it is natural to assume $\lambda_{1} 
\geq \lambda_{2}$.  Then in the vicinity of $ x \simeq 0.5\%$, where d-wave
superconductivity begins to appear, we can assume $\Delta_{2}/\Delta_{1}
\ll 1$.  Note that $\mu \sim 2300(x-0.0675)$ K \cite{66} in the whole region.  Then combining
Eqs. (15) and (16) we find
\bea
\lambda_{2}^{-1}-\lambda_{1}^{-1} & \simeq & 4 \pi T \mu^{2} \sum_{n}
\left\langle \frac{f^{2}}{[\omega_{n}^{2}+\Delta^{2}f^{2}]^{\frac{3}{2}}}
\right\rangle \\
& \simeq & \frac{2\mu^{2}}{\Delta^{2}(T)}(1 - 2(\ln 2)\frac{T}{\Delta_{0}})
\eea
where $\Delta^{2}(T) = \Delta_{1}^{2}(T) + \Delta_{2}^{2}(T)$.
Then Eq.(20) is solved as
\bea
(\frac{\Delta_{2}(T)}{\Delta_{2}(0)})^{2} & \simeq & 1 - 2(\ln 2)\frac{T}
{\Delta(0)}(\frac{\Delta_{1}(0)}{\Delta_{2}(0)})^{2}
\eea
or
\bea
T_{c2} & \simeq & \frac{1}{2 (\ln 2)} \frac{\Delta_{2}^{2}(0)}{\Delta(0)}
\eea
On the other hand the superfluid density in the gossamer superconductivity
is given by \cite{44}
\bea
\rho_{s}(T) &=& 2 \pi T \Delta_{2}^{2}(T) \sum_{n} Re \left\langle
\frac{f^{2}}{[(\sqrt{\omega_{n}^{2}+\Delta_{2}^{2}f^{2}}-i\mu)^{2}
+\Delta_{1}^{2}f^{2}]^{\frac{3}{2}}} \right\rangle
\eea
which gives 
\bea
\rho_{s}(0) & \simeq & \frac{\Delta_{2}^{2}(0)}{\Delta^{2}(0)}
\eea
Finally we find
\bea
T_{c2} &=& \frac{1}{2(\ln 2)}\Delta(0)\rho_{s}(0)
\eea
Since $\lambda^{-2}(0) = \frac{4\pi e^{2}}{m^{*}}n \rho_{s}(0)$ the
above relation can be interpreted as the celebrated Uemura relation \cite{62},
which could not be obtained within the framework of the BCS theory.

Also the present phase diagram suggests that the optimally doped superconductor
sits at the boundary of dDW I and dDW II.  Of course the present analysis
requires further elaboration.  Nevertheless, the present model appears
to describe qualitatively the phase diagram of high-T$_{c}$ cuprate
superconductors.  Also, Fig. 5 suggests naturally that both dDW III and
d-wave superconductivity terminate at $\mu/\Delta_{01} = 1.06$ (or x =25\%),
implying the quantum critical point (QCP) at x= 25\%.  We have mentioned
previously that the d-wave superconductivity in CeCoIn$_{5}$ is also most
likely ``gossamer''.

\section{Conclusions}

We have seen previously that most of the metallic ground states in
high-T$_{c}$ cuprates, heavy-fermion conductors and organic conductors
belong to one of the mean field ground states: a) unconventional 
superconductivity, b) unconventional density wave; or c) the coexistence
of both unconventional superconductivity and UDW.  The present study suggests
that perhaps a) most of the pseudogap phase or ``non-Fermi liquid'' behaviors
are related to UDW; and b) the superconducting 
phases in both high-T$_{c}$ cuprates
and CeCoIn$_{5}$ are gossamer; and c) the superconductivity in 
$\kappa$-(ET)$_{2}$ salts and in Bechgaard salts (TMTSF)$_{2}$PF$_{6}$ also
appear to be gossamer \cite{63,64}.  In the last system the superconductivity
is expected to be triplet and should contain an 
unconventional spin density wave (USDW)
\cite{53,54}.  This suggests that there are a variety of gossamer
superconductors, which await our exploration.

\begin{acknowledgement}
First of all we thank Carmen Almasan, Tao Hu and Viorel Sandu for 
providing us the data we used in Fig. 3 and Fig. 4.  
During the course of this work we have benefitted 
from enlightening discussions with Carlo di Castro, Andreas Greco,
Peter Horsch, Walter Metzner, Alejandro Muramatsu, Maurice Rice, Manfred
Sigrist, Peter Thalmeier and Roland Zeyher.  HW acknowledges support from
the Korean Research Foundation (KRF) through Grant No. R05-2004-000-10814.
\end{acknowledgement}

\end{document}